\documentclass[]{aa}
\usepackage{graphicx}
\usepackage{txfonts}

\newcommand{\msun}{\mbox{${\rm M}_\odot$}}

\newcommand{\rsun}{\mbox{${\rm R}_\odot$}}

\begin{document}

\title{A microquasar shot out from its birth place}
\author{I.F. Mirabel\inst{1, 2}, I. Rodrigues\inst{1, 3}, and Q.Z. Liu\inst{1, 4}}

\institute{Service d'Astrophysique, CEA-Saclay, 91191 Gif-sur-Yvette, France
 \and
       Instituto de Astronom\'\i a y F\'\i sica del Espacio/Conicet. Bs As, Argentina
 \and
      Instituto de F\'\i sica, Universidade Federal do Rio Grande do Sul, C.P. 15001, 91501-970, Porto Alegre, RS, Brazil
 \and       Purple Mountain Observatory, Chinese Academy of Sciences, Nanjing 210008, P.R. China}

\date{Received 31 March 2004/ Accepted 25 May 2004 (Published in A\&A, 422, L29, 2004)}

\offprints{I.F. Mirabel (email: fmirabel@cea.fr)}

\abstract{We show that the microquasar  LS\,I\,+61$^{\circ}$ 303 is 
running away from its birth place in a young complex of massive stars.  
The supernova explosion that formed the compact object shot out the x-ray 
binary with a linear momentum of 430$\pm$140 M$_{\odot}$ km s$^{-1}$, which is comparable 
to the linear momenta found in solitary runaway neutron stars and millisecond pulsars. 
The properties of the binary system and its runaway motion of 27$\pm$6 km s$^{-1}$ 
imply that the natal supernova was asymmetric and that the upper limit for the mass 
that could have been suddenly ejected in the explosion is $\sim$2 M$_{\odot}$. 
The initial mass of the progenitor star of the compact object that is inferred depends on whether 
the formation of massive stars in the parent stellar cluster was coeval or a sequential process.    
%

      }

\authorrunning{I.F. Mirabel et al.}
\maketitle

\section{Introduction}


%




Neutron stars are known to have large transverse motions on the plane of the 
sky which are believed to result from natal kicks imparted by 
supernova explosions. But the distances of solitary neutron stars 
are usually uncertain and it is not possible to determine the motion along 
the line of sight. To constrain the strength of the natal kick 
imparted to a particular neutron star, the distance and actual runaway 
space velocity in three dimensions must be known.   
If a compact object is accompanied by a mass-donor star, it is possible 
to determine the radial velocity, proper motion, and distance of the 
system. Then one can derive the space velocity, track the path in the sky, 
and estimate the strength of the natal kick,  
after correcting for the peculiar streaming motion of the parent cluster 
of stars. 

This kinematic method together
with the development of astrometric facilities at radio (VLBI) and
optical (e.g. GAIA) wavelengths will provide in the future powerful
observational tests for models of the evolution of massive
stars in binary systems and collapsars. Presently, the 
most accurate proper motions of x-ray binaries can be obtained following 
at radio wavelengths with Very Long Baseline Interferometry (VLBI) the 
motion in the sky of compact microquasar jets, as done recently 
for some x-ray binaries (e.g. Mirabel et al. 2001; Rib\'o et al. 2002; 
Mirabel \& Rodrigues, 2003a,b).

\section{LS\,I\,$+61^\circ 303$ and its kinematics}

LS\,I\,$+61^\circ 303$ (Gregory \& Taylor 1978) is a high mass x-ray binary (HMXB),  classified as
a Be/x-ray system~(Gregory \& Taylor 1978; Bignami et al. 1981).  The compact object is 
a neutron star or black hole (Massi, 2004) of 2$\pm$1 M$_{\odot}$
orbiting around a 14$\pm$4 M$_{\odot}$ donor star in an eccentric orbit with a period of 26.5 days (see Table 1).
LS\,I\,$+61^\circ 303$ is of particular interest because it is located inside the 95\% confidence radius of an
unidentified high energy source observed with EGRET on board the Compton Gamma-Ray Observatory (Kniffen et al. 1997).
The variable radio counterpart of LS\,I\,$+61^\circ 303$ (Gregory et al. 1979) has been resolved at milliarcsecond scales 
as a rapidly precessing relativistic compact jet  (Massi et al. 2001, 2004).

\begin{table*}[t]
\begin{center}
\caption{Basic data on the microquasar LS\,I\,$+61^\circ 303$ and the cluster IC\,1805.}
\begin{tabular}{llcccc}
\hline \hline \noalign{\smallskip}
                        &        & ~~~~~~~~~~~~LS\,I\,$+61^\circ 303$ &              & ~~~~~~~~~~IC\,1805  & \\
 \noalign{\smallskip} \hline \noalign{\smallskip}
 l                               & [$^\circ$]      &  135.68                & (1)       & 134.73          & (2) \\
 \noalign{\smallskip} \hline \noalign{\smallskip}
 b                               & [$^\circ$]      &  +1.09                 & (1)       & +0.92           & (2) \\
 \noalign{\smallskip} \hline \noalign{\smallskip}
 $\mu_{\alpha } \cos{\delta}$    & [mas yr$^{-1}$] &  0.97 $\pm$ 0.26       & (1)       & $-1.02 \pm$ 0.4 &   (2)  \\
 \noalign{\smallskip} \hline \noalign{\smallskip}
 $\mu_{\delta}$                  & [mas yr$^{-1}$] & $-1.21 \pm$ 0.3        & (1)       & $-0.88 \pm$ 0.4 &   (2)  \\
 \noalign{\smallskip} \hline \noalign{\smallskip}
 D                               & [kpc]           & 2.3  $\pm$ 0.4         & (3, 4)    & 2.3   $\pm$ 0.1 &  (5)  \\
 \noalign{\smallskip} \hline \noalign{\smallskip}
 V$_\mathrm{helio}$                     &  [km s$^{-1}$]  & $-55  \pm$ 4           & (6)       & $-41.2 \pm$ 3   & (2)  \\
 \noalign{\smallskip} \hline \noalign{\smallskip}
 U                               &  [km s$^{-1}$]  & 37.7 $\pm$ 4.1         &           & 42.9 $\pm$ 4.4  &    \\
 V                               &  [km s$^{-1}$]  & $-37.0 \pm$ 4.1        &           & $-11.8 \pm$ 4.5 &     \\
 W                               &  [km s$^{-1}$]  & $-1.8  \pm$ 3.4        &           & $-6.9  \pm$ 4.0 &    \\
 \noalign{\smallskip} \hline \noalign{\smallskip}
 Lifetime/Age                    & [Myr]           & $<$ 20                 &           &  $<$5           & (5, 7) \\
 \noalign{\smallskip} \hline \noalign{\smallskip}
 M$_\mathrm{x}$                           & [\msun]         & 2 $\pm$ 1              & (6, 8)    &  --             &\\
 \noalign{\smallskip} \hline \noalign{\smallskip}
 M$_\mathrm{Be}$                        & [\msun]         & 14 $\pm$ 4             & (6, 8)    &  --             &\\
 \noalign{\smallskip} \hline \noalign{\smallskip}
 Spect. Type                     &                 & B0Ve                   & (5, 6, 9) &  --            &\\
 \noalign{\smallskip} \hline \noalign{\smallskip}
 Bin. eccentr.                   &                 & 0.7 $\pm$ 0.1        & (8, 12)      &  --            &\\
 \noalign{\smallskip} \hline \noalign{\smallskip}
 Binary mean sep.                &    [\rsun]      & 46                     & (10)      &  --            &\\
 \noalign{\smallskip} \hline \noalign{\smallskip}
 Binary period                   &  [days]         & 26.5                   & (10, 11)  &  --            &\\
\noalign{\smallskip} \hline

\end{tabular}
\end{center}
\begin{list}{}{}
\item{Note to Table 1: (1) Lestrade et al. 1999; (2) Dambis et al. 2001; (3) Frail \& Hjellming 1991; 
(4) Steele et al. 1998; (5) Massey et al. 1995; (6) Hutchings
\& Crampton 1981; (7) Dennison et al. 1997; (8) Mart{\'\i}
\& Paredes 1995; (9) Paredes \& Figueras 1986; (10)
Taylor et al. 1992; (11) Gregory 2002; (12) Casares et al. 2004.} 
\end{list} 
\end{table*}

 \begin{figure*}
 \begin{center}
\resizebox{1.0\hsize}{!}{\includegraphics{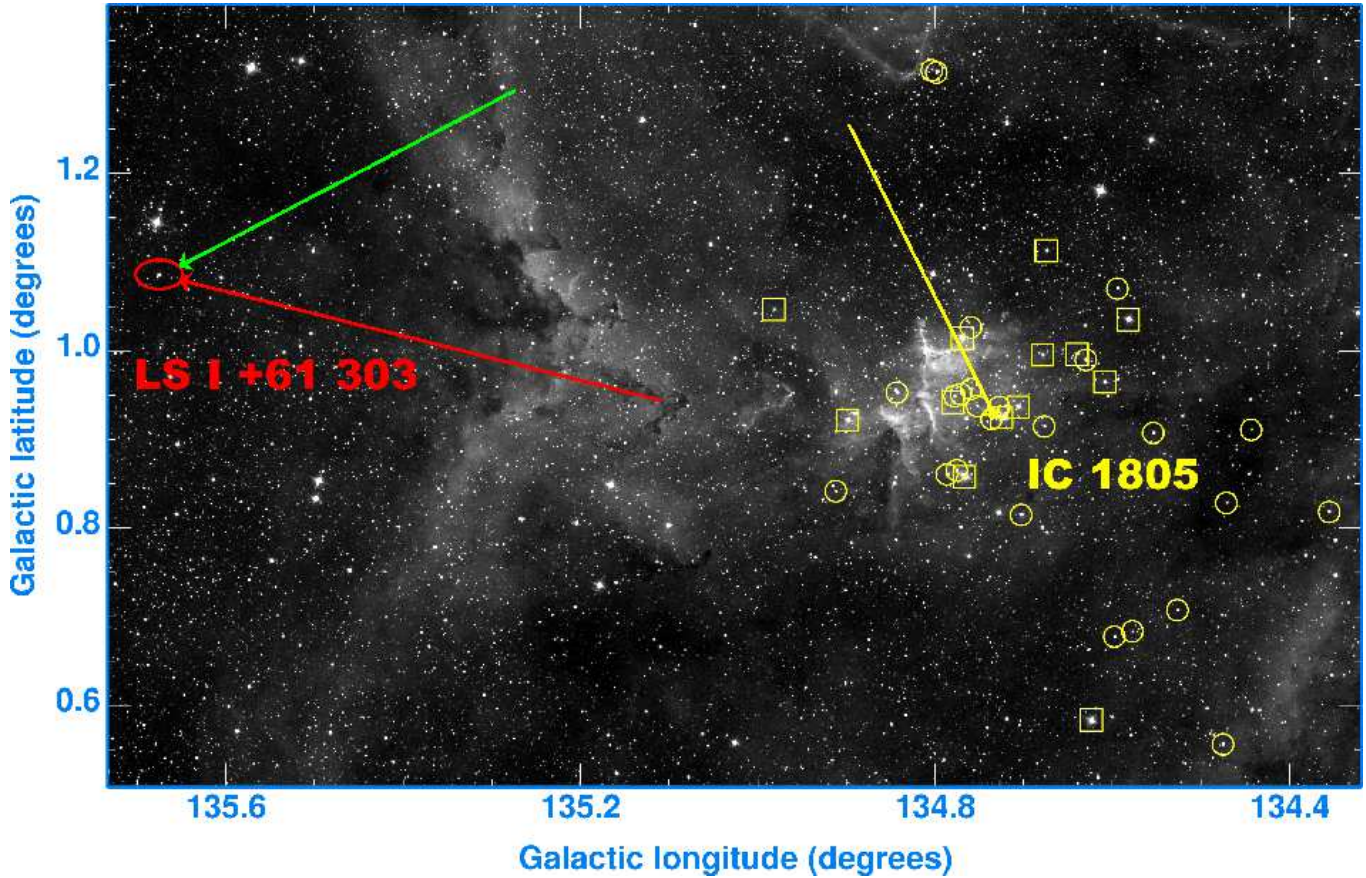}}
 \caption{The microquasar LS I $+61^\circ 303$ shot out at a velocity 
of 27$\pm$6 km s$^{-1}$ from the
stellar  cluster IC 1805 by the kick imparted to the compact object  
in a natal asymmetric supernova explosion. 
In this optical image (DSS, R
 filter) of the sky, LS I $+61^\circ 303$ is shown inside the red
 ellipse, and the confirmed member stars of IC 1805 with masses in the
 range of 7$-$85 M$_{\odot}$~ (Massey et al. 1995; Shi \& Hu 1999) 
are shown inside yellow symbols. The green
 arrow represents the motion in the sky of the radio counterpart of
 LS\,I\,$+61^\circ
303$ for the last 1 million years. The yellow arrow represents the
average Hipparcos motion of the 13 brightest stars of the cluster in
the Hipparcos catalog~(Dambis et al. 2001), which are
shown here inside yellow squares. The red arrow shows the motion of
the x-ray binary relative to the cluster of massive stars for the last
1 million years.}
  \label{fig sky}%
\end{center}  \end{figure*}

The motion in the
sky of the x-ray binary has been determined with high precision by VLBI 
astrometry of the compact jet (Lestrade et al. 1999).
Figure 1 shows that LS\,I\,$+61^\circ 303$ is near the cluster of massive 
stars and HII region IC 1805, which is part of the Cas OB6 association. 
As indicated in Table 1, the
estimated distances of the x-ray source and the star cluster are the same, 
despite the use of different observational
techniques. An independent indication of the cluster membership of 
LS\,I\,$+61^\circ 303$ is that the apparent
magnitude and colors of the donor star are consistent with those of the 
massive stellar members of the cluster.  That
is, the apparent magnitude and color of the donor star place it on the 
main sequence in the Hertzsprung-Russell (HR) diagram
of IC 1805.  Moreover, the x-ray binary and the stellar cluster are both 
falling towards the Galactic plane. Therefore,
LS\,I\,$+61^\circ 303$ may have been born in the cluster IC 1805.

At a distance of 2.3 kpc from the sun (Frail \& Hjellming 1991; Steele
et al. 1998) the x-ray binary is at a projected distance on the sky of
$\sim40$ pc from the center of IC 1805, and it is moving away from the
cluster with a relative spatial speed of $27\pm6$ km s$^{-1}$. Given
the relative transverse motion and using the angular extent of the
cluster it is inferred that the x-ray binary could have been ejected
from the cluster by a supernova explosion 1.7$\pm$0.7 ~Myr ago.  From
the mass of the x-ray binary and its space velocity, it is computed
that the energetic trigger imparted LS\,I\,$+61^\circ 303$ with a
runaway linear momentum of 430$\pm$140 M$_{\odot}$ km s$^{-1}$, which
is a typical value for runaway neutron star binaries (Toscano et
al. 1999).  

\section{The asymmetric natal supernova}

Following the analysis by Brandt \& Podsiadlowski (1995) and Nelemans et al. (1999) 
for symmetric mass ejection in the formation
of compact objects, and from the properties of LS\,I\,$+61^\circ 303$ (Table 1)
we estimate the maximum amount of mass that could have been suddenly ejected to
accelerate the binary to a runaway speed of $27\pm6$
km\,s$^{-1}$.    Assuming that the binary period and eccentricity did not
change considerably since the SN event, and using equations (5) and (7) of 
Nelemans et al. (1999) it is found that the mass that could have been  
suddenly ejected in a spherically symmetric SN explosion (which is the maximum possible) 
is $\Delta$M$_\mathrm{SN}$ =  2.0$\pm$1.0~\msun. 
As expected for a SN that took place
$\geq$$1$~Myr ago, no associated SN remnant in the X-rays and/or
radio wavelengths is found.

Van den Heuvel et al. (2000) have shown that a sudden spherically symmetric 
mass loss $\Delta$M$_\mathrm{SN}$ from a binary system of total mass 
M$_\mathrm{sys}$ would induce an orbital eccentricity e = $\Delta$M$_\mathrm{SN}$ $\times$ M$_\mathrm{sys}$$^{-1}$. 
Since $\Delta$M$_\mathrm{SN}$ = 2.0$\pm$1.0~\msun and M$_\mathrm{sys}$ = 16$\pm$4~\msun, this implies 
that a symmetric explosion would have produced an orbital eccentricity e = 0.12, 
which is much lower than the actual eccentricity  e = 0.7$\pm$0.1 (Table 1). 
Therefore, the SN explosion was asymmetric.
The asymmetric natal explosion imparted a kick to the compact object, 
which caused a smaller runaway velocity of the system V$_\mathrm{run}$ = 27$\pm$6 km s$^{-1}$, 
since the compact object had to drag its companion of 14$\pm$4~\msun. 


\section{Birth place in the cluster IC 1805}

More massive stars evolve to the SN stage faster, and the initial mass of 
the progenitor of the compact object in
LS\,I\,$+61^\circ 303$ could in principle  be estimated assuming that it 
was contemporary of the massive stars presently being
observed in the cluster. IC 1805 is one of the best studied clusters in the Milky Way 
(Guetter \& Vrba 1989, Sung \& Lee 1995; Massey et al. 1995; 
Shi \& Hu 1999). It contains 10 main sequence stars 
with masses in the range of
20$-$85 M$_{\odot}$ that were formed in the last $1-3 \times 10^6$ years. 
However, the HR diagram of Guetter \& Vrba (1989) and Massey et al. (1995) leave room 
for the possibility of an older population of stars in IC 1805, with ages of 10$-$20 Myr.  
Since it is not clear whether these stars belong to IC 1805 or to a foreground population,  
at present the initial mass and age of the progenitor of the neutron star is uncertain.   
 

If the cluster was formed in the last $4-5 \times 10^6$ years and LS\,I\,$+61^\circ 303$ was ejected from it $\geq$ 
1 Myr ago, the progenitor of the compact object had a lifetime $\leq$ 4 Myrs before it exploded as a SN. From
current models of stellar evolution (Schaller et al. 1992) it is inferred that only stars of $\geq$ 60 M$_{\odot}$ have
such short lifetimes before collapse. This would be a lower limit for the progenitor mass 
since stars with masses of up to 85 M$_{\odot}$ are
presently found in IC 1805 (Sung \& Lee 1995; Massey et al. 1995). The masses of the donor
and compact object are 14$\pm$4~M$_{\odot}$ and 2$\pm$1~M$_{\odot}$ respectively, 
and in the explosion not more than 
$\sim$2 M$_{\odot}$ were ejected. Therefore, if the progenitor was 
contemporary of the massive stars in IC 1805,  before collapse it must have 
lost $\geq$ 50 M$_{\odot}$, that is $\geq$ 90\% of its initial mass.


We point out that the two best studied soft gamma-ray repeaters, 
which are young neutron stars, are located in dust enshrouded clusters of massive stars 
(Mirabel et. al 2000). But unless it is known whether the massive star formation 
in these clusters was coeval or a continuous process, we don't know whether stars with masses 
$\geq$ 60 M$_{\odot}$ can end as neutron stars and black holes of low mass.

On the other hand, in the context of binary evolution models (Ergma \& van den Heuvel, 1998),  
the progenitor of the compact object in LS\,I\,$+61^\circ 303$ would be much older, say 10 million years, 
as its companion has a mass of about 14 M$_{\odot}$, and mass transfer between the components 
of the binary may have increased its mass. So the progenitor star might have had a smaller mass, 
say $15-20$  M$_{\odot}$. 



\section{Conclusion}

The compact object in LS\,I\,$+61^\circ 303$ was striken by an asymmetric 
natal supernova dragging the binary system with a
runaway linear momentum of 430$\pm$140 M$_{\odot}$ km s$^{-1}$. Large
runaway linear momenta comparable to those of solitary neutron stars
and millisecond pulsars were reported for the neutron star x-ray
binary LS 5039 (Rib\'o et al. 2002) and GRO J1655-40, which contains a
black hole of $\leq$ 7 M$_{\odot}$ (Mirabel et al. 2002). On the
contrary, Cygnus X-1 which contains a black hole of $\sim$10
M$_{\odot}$ was formed in situ and did not receive an energetic
trigger from a supernova (Mirabel \& Rodrigues, 2003a).  Although the
number statistics is still low, these preliminary results are
consistent with evolutionary models for binary massive stars 
(Balberg \& Shapiro 2001; Fryer et
al. 2002), where neutron stars and low-mass black holes form in
energetic supernova explosions, whereas the black holes with the
larger masses form in underluminous supernovae or even in complete darkness.




\acknowledgements We thank the referee Gijs Nelemans for important comments that helped to improve our interpretation 
of the observations. We also thank J. Casares and collaborators for communicating to us the determination 
of the orbital eccentricity prior to publication.  
We have profit from discussions with I. Negueruela, Ph. Podsiadlowski, M. Rib\'o, V. Kalogera, P. Massey, 
R. Henriksen, V. Niemela and D.R. Gies. I.R. acknowledges support from a post-doctoral fellowship of
the Conselho Nacional de Desenvolvimento Cient\'\i fico e Tecnol\'ogico (CNPq) of Brazil. QZL is partially supported by
973 Project through Grant G1999075405 and NSFC through Grant 10173026.

\end{document}